\documentclass[cameraready]{Interspeech}

\usepackage{placeins}
\usepackage{microtype}
\usepackage{stfloats}
\usepackage{url}

\newcommand{\tcell}[1]{\begin{tabular}[t]{@{}l@{}}#1\end{tabular}}

\title{MOV-AAD: A Large-Scale Multimodal Dataset for Auditory Attention Decoding During Moving Conversations}

\author[affiliation={1,2}, orcid=0009-0008-0625-8711, equalcontribution]{Xiaomin}{He}
\author[affiliation={1,2}, orcid=0009-0000-5486-5913, equalcontribution]{Vishal}{Choudhari}
\author[affiliation={1}]{Tristan J.}{Spratt}
\author[affiliation={1}]{Aarya}{Raghavan}
\author[affiliation={1}]{Richard T.}{Lee}
\author[affiliation={1,2},orcid = 0000-0002-2987-759X,correspondingauthor]{Nima}{Mesgarani}

\address{
    $^1$ Department of Electrical Engineering, Columbia University, New York, NY, USA \\
    $^2$ Zuckerman Mind Brain Behavior Institute, Columbia University, New York, NY, USA
}
\email{xh2369@columbia.edu, vishal.c@columbia.edu, tjs2198@columbia.edu, ar4634@columbia.edu, rtl2118@columbia.edu,nima@ee.columbia.edu}


\usepackage{comment}
\usepackage{caption}

\usepackage{booktabs}
\usepackage{tabularx}
\usepackage{array}

\usepackage{makecell}

\begin{document}

\maketitle

\begin{abstract}
   Auditory attention decoding (AAD) is often evaluated on static, simplified speech scenes that poorly match everyday listening. We introduce MOV-AAD, a large-scale dataset for studying auditory attention under moving, naturalistic conversations. MOV-AAD combines 64-channel EEG with synchronized physiological recordings, including eye tracking, respiration, galvanic skin response, heart rate, peripheral oxygen saturation, body temperature, body motion, and photoplethysmography, enabling analysis of cross-modal neural and physiological markers of attention and listening effort. This dataset uses a more ecologically valid listening paradigm with dynamically moving conversational speech sources and behavioral measures of attentional engagement. MOV-AAD supports research on robust AAD in realistic spatial dynamics, multimodal attention modeling, listening effort, and intersubject neural responses, providing a resource for benchmarking selective auditory attention in naturalistic listening.



\end{abstract}

\section{Introduction}

Selective auditory attention enables listeners to focus on a target speaker in noisy environments, commonly known as the “cocktail party effect” \cite{cherry_experiments_1953}. Neural studies using EEG and MEG have demonstrated that cortical activity preferentially tracks the speech envelope of the attended speaker \cite{ding_emergence_2012,ziongolumbic_mechanisms_2013,osullivan_attentional_2015}, forming the basis of auditory attention decoding (AAD) in multi-speaker scenarios. These findings have motivated research on neural decoding algorithms and neuro-steered hearing technologies.

Attentive listening, however, is not solely a cortical phenomenon. Selective attention and listening effort engage distributed neural and autonomic systems. Peripheral physiological markers such as pupillometry, heart rate variability, respiration dynamics, and electrodermal activity (e.g. galvanic skin response) have been linked to cognitive load, arousal regulation, and sustained attention during speech processing \cite{zekveld_cognitive_2014,seeman_comparison_2015,mackersie_subjective_2011, rochet-capellan_changes_2013}. Eye-tracking measures including pupil dilation and gaze stability similarly reflect task engagement and top-down control processes \cite{kang_pupil_2014,he_neural_2025,eckstein_beyond_2016}. Intersubject correlation (ISC) analyses further reveal shared stimulus-driven neural dynamics during attentive engagement \cite{dmochowski_correlated_2012, cohen_memorable_2016,holtze_ear-eeg_2022}. These findings suggest that auditory attention involves coordinated interactions across neural and peripheral physiological systems.

However, most publicly available AAD datasets focus primarily on EEG and lack synchronized autonomic and behavioral measurements. Moreover, many existing datasets rely on spatially stationary speakers, synthetic mixtures, or constrained acoustic paradigms that only partially approximate real-world listening environments \cite{debener_neural_2024, fuglsang_eeg_2018}. The lack of naturalistic multimodal recordings limits investigations of how neural and physiological processes jointly support attention and effort in dynamic acoustic scenes.

Here we introduce \textbf{MOV-AAD}\footnote{Available at   \url{https://github.com/naplab/MOV-AAD}}, a large-scale multimodal dataset designed to support research on auditory attention decoding and cognitive state assessment under naturalistic listening conditions. The dataset integrates 64-channel EEG with synchronized physiological and behavioral modalities, including heart rate, respiration flow and effort, galvanic skin response (GSR), peripheral oxygen saturation (SpO$_2$), body temperature, body motion (accelerometry), photoplethysmography (PPG), and binocular eye tracking (pupil dilation and gaze location), recorded from 50 participants. By combining neural and autonomic measures under dynamic listening conditions, MOV-AAD provides a foundation for investigating how top-down attention is coordinated across cortical and peripheral systems.

To position MOV-AAD within the existing literature, Table~\ref{tab:aad_datasets} summarizes representative publicly available non-invasive AAD datasets. While datasets such as DTU \cite{fuglsang_eeg_2018, fuglsang_noise-robust_2017} and KULeuven \cite{das_auditory_2019,biesmans_auditory-inspired_2017} provide high-density EEG recordings in two-speaker paradigms, most employ spatially stationary speech mixtures and lack synchronized peripheral physiological measurements. The PhyAAt dataset \cite{bajaj_phyaat_2020} incorporates selected physiological signals but primarily focuses on single-speaker listening with noise rather than competitive multi-talker attention. Compared to these resources, MOV-AAD uniquely combines (i) dynamic spatial movement of conversational speech sources with turn-taking, (ii) a substantial participant cohort (N=50), and (iii) comprehensive multimodal recordings integrating EEG with autonomic and behavioral measures.


\begin{table*}[!t]
\centering
\footnotesize
\setlength{\tabcolsep}{3pt}
\renewcommand{\arraystretch}{1.12}
\caption{\textbf{Comparison of representative non-invasive AAD datasets.} \textup{MOV-AAD uniquely combines dynamic spatial movement, conversational speech, and synchronized multimodal physiological recordings under ecologically realistic conditions.}}
\label{tab:aad_datasets}

\begin{tabularx}{\textwidth}{
@{}
>{\raggedright\arraybackslash}p{1.75cm}
>{\centering\arraybackslash}p{0.5cm}
>{\centering\arraybackslash}p{0.5cm}
>{\centering\arraybackslash}p{1.0cm}
>{\raggedright\arraybackslash}p{2.9cm}
>{\raggedright\arraybackslash}p{3.8 cm}
>{\raggedright\arraybackslash}p{2.6 cm}
>{\raggedright\arraybackslash}p{3.4cm}
@{}
}
\toprule
\textbf{Dataset} & \textbf{Year} & \textbf{Sub.} & \textbf{EEG ch.}
& \textbf{Other modalities}
& \textbf{Speech type}
& \textbf{Background noise}
& \textbf{Spatial setup} \\
\midrule

KULeuven \cite{biesmans_auditory-inspired_2017}
& 2017 & 16 & 64
& --
& Continuous speech, overlapping two-talker
& --
& Static at $\pm90^\circ$ \\

DTU \cite{fuglsang_noise-robust_2017}
& 2017 & 18 & 64
& EOG
& \tcell{Continuous speech, single-talker\\and overlapping two-talker}
& --
& Static at $\pm60^\circ$ \\

PhyAAt \cite{bajaj_phyaat_2020}
& 2020 & 25 & 14
& PPG, GSR
& Sentences, single-talker
& \tcell{Unspecified noise;\\varied SNRs}
& -- \\

NJU AAD \cite{zhang_nju_nodate}
& 2023 & 21 & 32
& --
& Continuous speech, overlapping two-talker
& --
& \tcell{Static; random left-right\\pair from 14 azimuths} \\

SparrKULee \cite{accou_sparrkulee_2024}
& 2024 & \textbf{85} & 64
& --
& Continuous speech, single-talker
& \tcell{Speech-weighted noise\\in subset, -5 dB}
& -- \\

\textbf{MOV-AAD}
& \textbf{2026} & 50 & 64
& \textbf{\tcell{PPG, GSR, eye tracking,\\respiration, SpO$_2$,\\temperature, motion}}
& \textbf{\tcell{Conversations with turn-taking,\\single-talker and overlapping \\two-talker}}
& \tcell{Babble and street-scene \\ noise, -9 / -12 dB }
& \textbf{\tcell{Moving sources\\within $\pm90^\circ$}} \\

\bottomrule
\end{tabularx}
\end{table*}

Each participant completed both single-conversation and multi-conversation tasks. In the single-conversation condition (40 trials), participants listened to conversational narratives presented from a spatially moving source. In the multi-conversation condition (56 trials), they attended to one speaker among dynamically moving competing conversational streams spanning $\pm90^\circ$ azimuth. To behaviorally track attention, participants performed a repeated-word detection task during listening, which has been shown to correlate with attentional engagement and task difficulty \cite{marinato_object-based_2019,laffere_effects_2020}. The inclusion of spatial movement and naturalistic dialogue enhances ecological validity and approximates the complexity of real-world auditory scenes \cite{bednar_where_2020,choudhari_brain-controlled_nodate}.

Across all tasks, synchronized 64-channel EEG and multimodal physiological signals were recorded from 50 participants, comprising 4,800 attentive-listening trials. As one of the largest publicly available multimodal resources for non-invasive auditory attention research, MOV-AAD enables dynamic-condition benchmarking, listening-effort analysis, and within- and cross-subject modeling.

\section{Dataset Description}

\subsection{Participants}

The study protocol was approved by the Institutional Review Board of the authors' university. Fifty participants (28 male at birth; age 18-38 years, mean $=24$, SD $=4.5$) with self-reported normal hearing and no history of neurological or attention disorders were recruited between October 2022 and June 2025. Written informed consent was obtained prior to participation, and all participants were compensated.

\subsection{Speech Materials and Spatialization}

Speech stimuli were constructed as naturalistic conversational streams spoken by native American English talkers. Each conversation consisted of temporally continuous speech segments with occasional embedded repeated words used for attention monitoring. The mean interval between repeated-word onsets within a stream was 7.0 s (SD = 1.0 s).

All speech streams were spatialized using head-related transfer functions (HRTFs) and presented through earphones. During moving-speaker conditions, conversations moved continuously within the frontal-horizontal plane (azimuth range: -$90^\circ$ to +$90^\circ$). The perceived distance of each stream was held constant across trials. In multi-conversation trials, streams were matched in root-mean-square (RMS) power.

Diotic background noise (pedestrian noise or speech babble) was added at levels 9 or 12 dB below the power of each conversation stream.

\begin{table}[th] 
\centering
\footnotesize
\setlength{\tabcolsep}{4pt}
\caption{\textbf{Dataset summary}. \textup{Key parameters of the MOV-AAD dataset, including modalities, tasks, and stimulus configuration.}}
\renewcommand{\arraystretch}{1.2}
\begin{tabular}{p{1.7cm} p{5.6cm}} 
\toprule
\textbf{Feature} & \textbf{Description} \\
\midrule
\textbf{Modalities} & 64-ch EEG, eye tracking (pupil, gaze), respiration (airflow, effort), GSR, PPG, SpO\textsubscript{2}, skin temperature, accelerometer, behavioral responses \\
\textbf{Behavioral} & Localization report, Repeated-word detection \\
\textbf{Sampling rate} & 1200 Hz \\
\textbf{Participants} & 50 subjects (age 24 $\pm$ 4.5 years), normal hearing \\
\textbf{Tasks \& trials} & Repeated sentence task (120 sentences)\newline  
Localization (54 sentences behavior report)\newline  
Single-conversation (40 trials)\newline 
Multi-conversation (56 trials) \\
\textbf{Spatial setup} & HRTF-based moving sources (-$90^\circ$ to +$90^\circ$)\newline  Speakers RMS-matched \\
\textbf{Noise} & Diotic (pedestrian/babble) at -$9$, -$12$ dB \\
\textbf{Trial duration} & Repeated sentence task (5.2 $\pm$ 0.8 s); \newline Conversation tasks (40 $\pm$ 2.0 s) \\
\bottomrule
\end{tabular}
\label{tab:dataset_summary}
\end{table}

\subsection{Experiment Paradigm}




\begin{figure*}[t]
\centering
\includegraphics[width=0.95\linewidth]{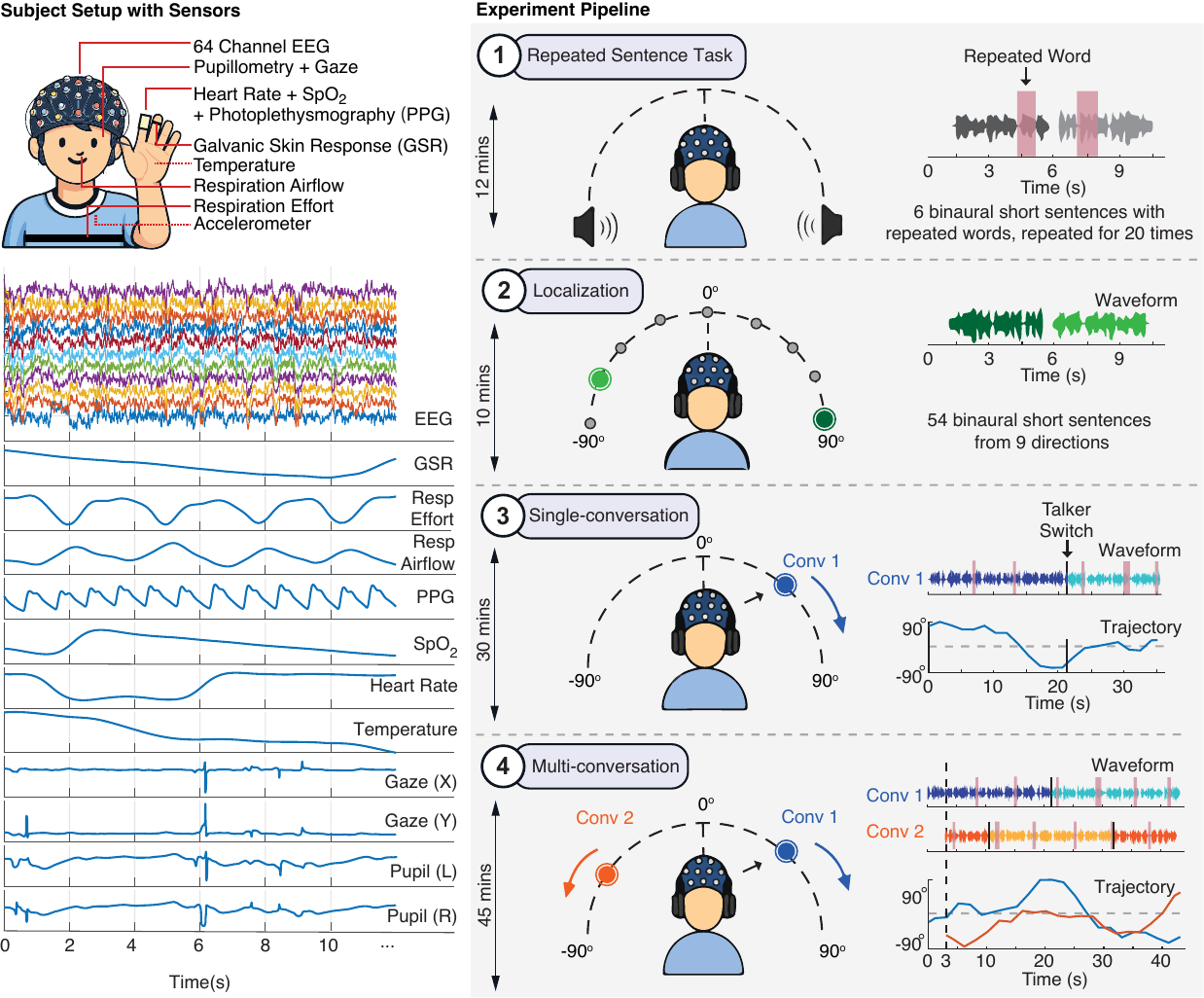}
\caption{\textbf{Overview of the MOV-AAD dataset and experimental paradigm.}
\textup{\textbf{Left:} Multimodal recording setup and example for synchronized signal traces.
\textbf{Right:} Full experimental pipeline and stimuli description.}}
\label{fig:dataset_overview}
\end{figure*}

\subsubsection{Repeated Sentence Task}
As an initial EEG reliability check, participants completed a repeated sentence task. Six short sentences, spoken by three male and three female talkers, were presented diotically without background noise and repeated 20 times, with sentence order shuffled across repetitions. Participants were instructed to listen attentively and press the buzzer when they detected one repeated word in each sentence. This task supported split-half or odd-even reliability analyses of neural responses to repeated speech stimuli, following prior work using repeated natural speech presentations as a sanity check for EEG data quality and response reliability \cite{hensch_reliability_2008}. The resulting recordings help evaluate whether reliable speech-evoked neural responses and stimulus-locked alignment could be observed within each participant.

\subsubsection{Localization Task}

Before the formal experiment, a localization task was conducted to familiarize participants with the spatial cues and to assess their spatial perception accuracy. 

Participants listened to 54 short spatialized sentences presented without background noise. Each sentence was rendered at one of nine evenly spaced azimuth locations ranging from -$90^\circ$ to +$90^\circ$ (i.e., -$90^\circ$, -$67.5^\circ$, -$45^\circ$, -$22.5^\circ$, $0^\circ$, $22.5^\circ$, $45^\circ$, $67.5^\circ$, $90^\circ$). After each presentation, participants reported the perceived direction of the sound source using a mouse-click response on a graphical interface. 

\subsubsection{Single Moving Speaker Conversation (SC)}

Participants completed 40 trials in which a single conversational stream moved continuously within the frontal horizontal plane (-90$^\circ$ to +90$^\circ$). Speaker identity varied within trials, and participants responded to embedded repeated words in a 1-back detection task.

\subsubsection{Multiple Moving Speakers Conversation (MC)}

This task extended the SC framework by presenting two independent conversational streams simultaneously. Participants completed 56 trials in which two spatially separated conversational streams moving within the same azimuth range. Participants were instructed to attend to the conversation that began first and to press a button upon detecting repeated words in the attended stream. The unattended stream began 3 s after the onset of the attended stream. In the attended conversation, one talker switch occurred at approximately 50\% of the trial duration, whereas in the unattended stream, two switches occurred at approximately 25\% and 75\%. Assignment of male and female talkers was counterbalanced across trials. Repeated-word events did not overlap across streams.

\subsection{Data Acquisition and Preprocessing}

EEG (64 channels) and peripheral physiological signals were acquired using g.tec sensors and a g.HIamp amplifier (g.tec medical engineering GmbH, Austria) at 1200 Hz with a 60 Hz notch filter. The peripheral modalities included GSR, triaxial accelerometry, respiratory flow (nasal airflow), respiratory effort (thoracic belt), SpO\textsubscript{2}, PPG, heart rate, and skin temperature. Eye-tracking signals, including pupil diameter and instantaneous horizontal and vertical gaze coordinates, were acquired using a Tobii Pro Nano eye tracker (Tobii AB, Sweden). All neural, physiological, and eye-tracking streams were synchronized and recorded through Simulink.

Behavioral responses (repeated-word detections and localization reports) were collected through a customized MATLAB-based interface and synchronized to physiological recordings. Audio playback was simultaneously recorded, and temporal alignment was established by cross-correlating the recorded audio with the original stimulus waveform.

The released dataset includes recordings from the repeated-sentence task 
($\sim$12 min per participant), SC tasks ($\sim$30 min per participant), and MC tasks ($\sim$45 min per participant); and behavioral reports from the localization task, SC and MC tasks. Specifically, for SC and MC trials, we provide EEG and peripheral signals at 1200 Hz together with aligned stimulus audio and repeated-word detection performance metrics at both the subject and trial levels. EEG channels with abnormal variance were automatically detected and replaced via spherical interpolation 
\cite{delorme_eeglab_2004, kang_distortions_2015, perrin_spherical_1989}. A 3 s 
pre-onset baseline segment is included for each trial. Peripheral modalities are 
released in minimally processed form, with 1200 Hz acquisition and hardware notch 
filtering only, without additional modality-specific preprocessing. For the localization task, trial- and subject-level localization metrics are provided.

\section{Method}
\subsection{Behavioral Validation Metrics}
\subsubsection{Localization Perception}

The localization task included nine discrete azimuth locations spanning -90° to +90°. For each participant, localization performance was quantified by accuracy and error. Localization accuracy was computed as the proportion of correctly identified source locations across trials, with chance performance defined as 1/9. 
Localization performance was further quantified using the mean absolute error (MAE) between the reported and the ground-truth stimulus location. For each participant, MAE was computed as

\begin{equation}
\mathrm{MAE} = \frac{1}{N}\sum_{i=1}^{N} \left| \hat{\theta}_i - \theta_i \right|,
\end{equation}

where $\theta_i$ denotes the true azimuth angle of the stimulus in trial $i$, $\hat{\theta}_i$ is the participant's reported azimuth, and $N$ is the total number of trials. Smaller MAE values indicate higher localization accuracy. Chance-level MAE was estimated assuming random responses among the nine possible azimuth locations, resulting in an expected error of approximately $66.7^\circ$.

\subsubsection{Repeated-word Detection Metrics}

Repeated-word performance was used to assess sustained attention during the conversation tasks. A repeated word was considered correctly detected if a space-bar press occurred within 2 s of the onset of the repeated word in the attended stream. A space-bar press was counted as a false alarm if no attended repeated-word onset occurred in the preceding 2 s window.

For each participant, we computed hit rate, precision, and F1 score separately for the SC and MC conditions. 




Statistical comparisons were performed using two-tailed Wilcoxon signed-rank tests. F1 scores were compared between MC and SC conditions as well as two streams within the MC condition (Attended vs. Unattended). For all tests, we report $p$-values and rank-based effect sizes ($r = Z / \sqrt{N}$), where $Z$ denotes the standardized statistic from the Wilcoxon signed-rank test and $N$ is the number of non-zero paired differences.


\subsection{Physiological Data Quality Assessment}

We perform a data quality sanity check by extracting a set of basic physiological features from the MC session, which include heart rate, skin temperature, skin conductance level (SCL, extracted from GSR), non-specific skin conductance response rate (NS-SCR, extracted from GSR), respiration frequency, accelerometer magnitude, SpO$_2$, heart rate variability (HRV, extracted from PPG) features, and pupil dilation.

GSR was characterized using both tonic and phasic measures. The slow-changing tonic component (SCL) was computed by low-pass filtering and averaging the raw GSR signal to capture baseline level. The phasic component (NS-SCR rate) was computed as the number of spontaneous SCR events per minute.

Accelerometer signals were summarized using the magnitude of tri-axial acceleration, expressed in units of gravity (g), as a proxy for body movement during the task. Since this measure includes the static gravitational component, a baseline around 1 g is expected, and deviations from this value reflect changes in movement and/or orientation.

HRV metrics were derived from PPG signals by detecting pulse peaks and computing inter-beat intervals. From these intervals, standard HRV metrics were calculated: SDNN (standard deviation of NN intervals), reflecting overall variability; and RMSSD (root mean square of successive differences), reflecting parasympathetic activity.

All features were computed within fixed-length analysis windows (40~s) and averaged across trials and participants. The resulting distributions were compared against expected physiological ranges based on prior literature\cite{dawson_electrodermal_2007,munoz_validity_2015,bourdillon_minimal_2017,mackersie_effects_2015} and device specifications.

\subsection{Baseline Analyses}





\subsubsection{Envelope Reconstruction}

For each trial, the speech envelope was extracted using the magnitude of the analytic signal (Hilbert transform), followed by 4th-order Butterworth low-pass filtering at 8 Hz. This approach follows common practice in cortical speech tracking studies targeting slow amplitude modulations \cite{holtze_ear-eeg_2022,mirkovic_decoding_2015}.

To reconstruct the speech envelope from EEG, data were resampled from 1200 Hz to 100 Hz, with an embedded anti-aliasing filter. Signals were normalized and nonlinearly compressed using $5\tanh(x/5)$ to mitigate extreme outliers. EEG was then band-pass filtered to 1-8 Hz using a zero-phase FIR filter (order 256). This frequency band captures low-frequency neural tracking of the speech envelope widely reported in EEG and MEG literature \cite{osullivan_attentional_2015,pasley_reconstructing_2012}.

Envelope reconstruction was performed using a backward model implemented in the mTRF Toolbox v2.3 \cite{crosse_linear_2021}. Decoders were trained with time lags from 0 to 400 ms. Ridge parameters were logarithmically spaced from $10^{-2}$ to $10^{4}$ and selected via leave-one-trial-out cross-validation. Reconstruction performance was quantified as the Pearson correlation between the reconstructed and true envelopes on the held-out trial. Single-speaker and competing-speaker sessions were modeled and evaluated separately.

\subsubsection{Trajectory Reconstruction}

Spatial trajectories were defined as the time-varying azimuth positions of each conversational stream. Trajectories were generated during stimulus design following the procedure described in \cite{choudhari_brain-controlled_nodate}. Briefly, speaker positions were constrained to the frontal half of the horizontal plane (-90° to +90°) and were generated using a first-order Markov chain with 1° spatial resolution. The resulting sequences were smoothed and upsampled to 10 Hz prior to spatial rendering via HRTF-based processing. The final trajectories were further interpolated to 100 Hz to match the neural sampling rate.

Trajectory reconstruction was performed following the decoding framework described in \cite{bednar_where_2020}. EEG signals were resampled to 100 Hz and robustly normalized prior to band-specific filtering. For delta-band analysis, EEG was band-pass filtered between 0.05-2 Hz using a zero-phase Butterworth filter. For alpha-band analysis, EEG was band-pass filtered between 8-12 Hz, and the analytic amplitude was obtained using the Hilbert transform.

Reconstruction was implemented using the same backward modeling framework described in the Envelope Reconstruction section. Decoders were trained with time lags from 0 to 250 ms. Ridge parameters were logarithmically spaced from $10^{-2}$ to $10^{4}$ and selected via leave-one-trial-out cross-validation.

Reconstruction was quantified as the Pearson correlation between reconstructed and true trajectories. In the multi-conversation condition, attended and unattended trajectories were reconstructed separately, and the resulting correlation coefficients were used for subsequent attention decoding analysis.

\begin{figure}[!t]
  \centering
  \includegraphics[width=1.05\linewidth]{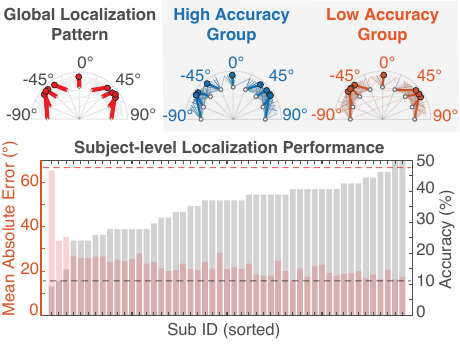}
  \caption{\textit{\textbf{Localization Perception.}}
\textup{\textbf{Top:} Localization patterns visualized in polar coordinates. For each subject, perceived angles were averaged for each true angle. Gray open circles denote true stimulus angles, and colored markers represent the across-subject average perceived angle for each true angle. Bottom panels compare high-accuracy (top 33\% of subjects by localization accuracy) and low-accuracy (bottom 33\%) groups, highlighting slight differences in localization profiles.
\textbf{Bottom:} Subject-level localization performance. 
Gray bars indicate localization accuracy, and red bars indicate mean absolute localization error. Subjects are sorted by accuracy. Dashed lines denote chance level (accuracy = $11.11\%$, black; mean absolute error $\approx 66.7^\circ$ for the -$90^\circ$ to $90^\circ$ range with nine response options, red).}}
  \label{fig:behavior_localization}
\end{figure}
\subsubsection{Auditory Attention Decoding (AAD)}

\begin{figure}[t]
  \centering
  \includegraphics[width=\linewidth]{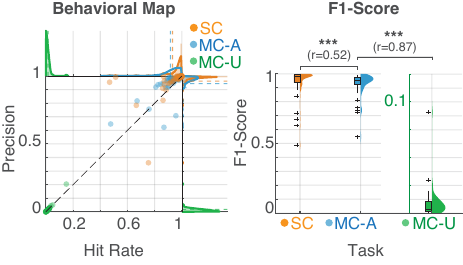}
  \caption{\textit{\textbf{Repeated-word detection performance in the conversation tasks. }} \textup{\textbf{Left: }behavioral map showing precision vs.\ hit rate for single-conversation (SC), multi-conversation attended (MC-A), and multi-conversation unattended (MC-U). \textbf{Right:} F1-score across tasks/streams. Asterisks denote significant paired differences (two-tailed Wilcoxon signed-rank test, $^{*}p<0.05$,$^{**}p<0.01$,$^{***}p<0.001$); the corresponding rank-based effect size $r$ is annotated.}}
  \label{fig:behavior_rw}
\end{figure}

\begin{figure*}[b]
\centering
\includegraphics[width=0.95\linewidth]{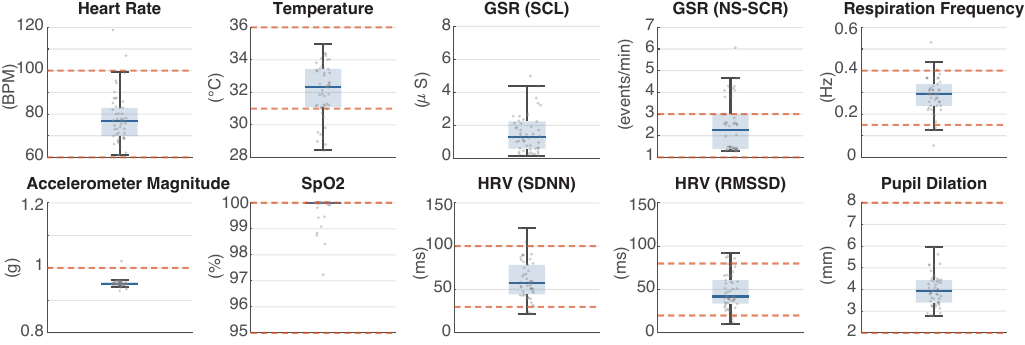}
\caption{
\textit{\textbf{Physiological data sanity check.}
\textup{Each panel shows distributions of physiological features (boxplots; median in blue). Gray dots represent participants. Orange dashed lines indicate expected physiological ranges based on prior literature and device specifications.}}
}
\label{fig:sanity_check}
\end{figure*}

For multi-conversation trials, AAD was performed separately for each participant using linear discriminant analysis (LDA) following \cite{bednar_where_2020}. For each trial, $r_{\mathrm{A}}$ and $r_{\mathrm{U}}$ denote the Pearson correlations between the reconstructed signal and the attended and unattended stimuli, respectively.

Trial-wise feature vectors were constructed as $[r_{\mathrm{A}}, r_{\mathrm{U}}]$. To form a balanced binary classification problem during training, swapped feature vectors $[r_{\mathrm{U}}, r_{\mathrm{A}}]$ were included as an alternative class.

Decoding was evaluated using leave-one-trial-out cross-validation across trials. In each fold, the held-out trial was tested using the true-order feature vector, while training was performed on both true-order and swapped-order samples from the remaining trials. Accuracy was defined as the proportion of correctly classified held-out true-order trials. Multiple feature setups were tested, including envelope-only, trajectory-only (delta or alpha), and combined feature sets.

Statistical significance was assessed using a permutation-based null distribution (1000 permutations). For each permutation, the identical cross-validation procedure was repeated with randomized training labels within each fold. Observed decoding accuracy was compared against the subject-specific null distribution and corrected for multiple comparisons.

\subsubsection{Intersubject Correlation (ISC)}
We quantified stimulus-locked response similarity across participants using ISC computed in correlated components of the multichannel signals \cite{cohen_memorable_2016, dmochowski_correlated_2012}. All modalities were notched-filtered at 60Hz and resampled to 100 Hz prior to analysis. Signal preparation was modality-specific:

\begin{itemize}
    \item \textbf{EEG:} Mean-centered across time (per channel) and   band-pass filtered using zero-phase FIR filters to either 1-8 Hz  (low-frequency band that manifests the envelope tracking) or 8-12 Hz (alpha band).

    \item \textbf{Eye tracking:} Pupil size and horizontal or vertical gaze instantaneous position were averaged across eyes after excluding samples marked invalid by the eye tracker (Tobii Pro Nano, Tobii AB, Sweden).

    \item \textbf{Respiration flow, effort, and PPG:} Detrended or high-pass filtered at 0.05 Hz to remove slow drifts.
    
    \item \textbf{Heart rate and SpO\textsubscript{2}:} High-pass filtered at 0.05 Hz and subsequently low-pass filtered at 1 Hz.
    
    \item \textbf{GSR:} Median filtered to reduce impulse artifacts.
\end{itemize}

For each trial, data were arranged as $\mathbf{X}(t)\in\mathbb{R}^{D\times N}$ (channels $\times$ subjects). Spatial filters were obtained by maximizing between-subject covariance relative to within-subject covariance using generalized eigenvalue decomposition. Within-subject covariance was shrinkage-regularized \cite{cohen_memorable_2016}. We retained three EEG components and one component for single-channel modalities.

Trial-wise ISC scores were computed by correlating each participant’s projected component time series with the average of the remaining participants. Trial-level ISC values were averaged across trials to yield one ISC value per subject. However, for slowly varying physiological measures (e.g., temperature and SpO$_2$), trial-wise ISC scores were unstable due to limited within-trial temporal variability. Therefore, trials were aggregated in non-overlapping blocks of 28 trials (half of the experiment) to compute ISC at a longer time scale.

Chance-level ISC was estimated using surrogate data generated by phase randomization, which preserves temporal autocorrelation while disrupting cross-subject alignment \cite{cohen_memorable_2016}. For each subject, significance $p$-values were computed by comparing the observed ISC against the permutation-derived null distribution (100 permutations). Group-level significance was assessed using paired right-tailed tests comparing observed ISC to chance across subjects (paired $t$-test when ISC–chance differences passed a Lilliefors normality test; otherwise Wilcoxon signed-rank test). To quantify prevalence, the number of subjects with permutation test $p<0.05$ was evaluated using a binomial test against the expected false-positive rate.


\section{Results and Discussion}
 To demonstrate the validity and analytical potential of the dataset, we report behavioral performance, stimulus reconstruction accuracy, auditory attention decoding accuracy, and intersubject correlation using the standard evaluation metrics described in the Method section.

\subsection{Behavioral Validation}

\subsubsection{Localization Performance}



We used the localization task to assess the feasibility of our moving-conversation tracking paradigm. As shown in Fig.~\ref{fig:behavior_localization}, the polar visualization illustrates the mapping between true and perceived angles. Although localization responses showed reduced discriminability at extreme lateral positions, the overall angular structure was preserved, and perceived angles remained largely aligned with the correct hemifield across both higher-accuracy (top 33\%) and lower-accuracy (bottom 33\%) groups.

The greater variability observed at lateral angles is consistent with established characteristics of human spatial hearing, where localization precision typically decreases toward extreme azimuths due to cue saturation of interaural time and level differences and increased ambiguity \cite{mills_minimum_1958}. Such effects may be further amplified when non-individualized HRTFs are employed \cite{li_measurement_2020}. The observed performance quantifies the magnitude of these limitations: The spatial response patterns remained systematically structured, indicating that spatial cues were preserved at a level sufficient to support moving-speaker tracking at the group level.

At the subject level, localization performance showed moderate variability across participants (Fig.~\ref{fig:behavior_localization}, bottom). Localization accuracy ranged broadly but remained above chance level (approximately 11.1\%) for most participants, while mean absolute localization error was substantially below the level expected from random guessing chance level (approximately $ 66.7^\circ$). These results indicate that although spatial perception accuracy varied across individuals, the majority of participants maintained reliable spatial discrimination under the current experimental setup.

\subsubsection{Repeated-word Detection Performance}


Repeated-word detection was conventionally used to evaluate participants' attentional engagement during tasks. As shown in Fig.~\ref{fig:behavior_rw}, repeated-word detection performance confirmed effective attentional engagement during the conversation tasks. 

In the behavioral map, SC and MC-A clustered in the high-precision, high-recall region for all participants, whereas MC-U showed markedly reduced detection performance. Consistently, F1-scores were significantly higher for MC-A compared to MC-U (Wilcoxon signed-rank test, $p<10^{-4}$, $r=0.87$), confirming successful attentional separation between attended and unattended streams. Although the difference between SC and MC-A was modest, F1-scores were also significantly higher in SC than in MC-A ($p<10^{-4}$, $r=0.52$), indicating slightly reduced detection performance when involving interfering conversations.


\subsection{Physiological Signal Sanity Check}

Basic physiological features, including heart rate, skin temperature, GSR (quantified using SCL and NS-SCR), respiration frequency, accelerometer magnitude, SpO$_2$, HRV (quantified using SDNN and RMSSD), and pupil dilation, were extracted during the MC task (see Method for details), as shown in Fig.~\ref{fig:sanity_check}. These features exhibited distributions consistent with expected physiological ranges defined based on prior literature and device specifications. Most measurements fell within or close to these ranges, indicating overall good signal quality. A small number of outliers were observed, likely reflecting natural variability or minor recording artifacts.

\subsection{Stimulus Tracking Performance and AAD accuracy}

\begin{figure}[!t]
  \centering
  \includegraphics[width=\linewidth]{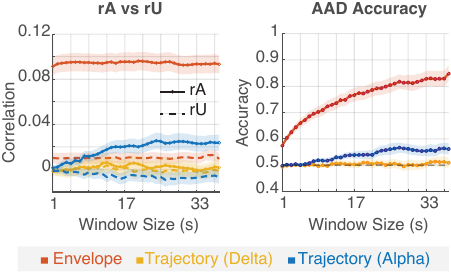}
  \caption{\textit{\textbf{Stimulus reconstruction correlation and AAD accuracy across window sizes.}}
  \textup{Sliding-window reconstruction was performed within each trial (hop size = 0.5~s). 
\textbf{  Left: } Reconstruction correlations for attended ($r_A$) and unattended ($r_U$) streams as a function of window size. Correlations are averaged across windows and trials per subject, then across subjects (mean $\pm$ 95\% CI). 
 \textbf{ Right:} AAD accuracy as a function of window size. For each subject and trial, accuracy was defined as the proportion of windows with $r_A > r_U$. Subject-level accuracy was obtained by averaging across trials, and the plotted curve shows the across-subject mean $\pm$ 95\% CI. Colors indicate feature type. Chance level (0.5) is indicated by the dashed line.}}
  \label{fig:aad_accuracy}
\end{figure}

\begin{figure}[!t]
  \centering
  \includegraphics[width=\linewidth]{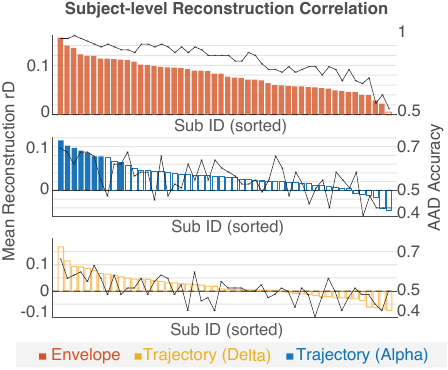}
\caption{\textit{\textbf{Subject-level stimulus reconstruction and AAD accuracy.}}
\textup{Bars show subject-level mean reconstruction correlation ($r_D$) averaged across trials for envelope or trajectory features. Subjects are sorted by their mean $r_D$. Filled bars denote subjects with $r_D$ significantly above chance (one-tailed Wilcoxon signed-rank test across trials on Fisher $z$-transformed correlations, Benjamini-Hochberg FDR-corrected $p_{\mathrm{FDR}}<0.05$); open bars indicate non-significant subjects. The black line (right axis) shows the corresponding full-trial AAD accuracy for each subject. The dashed horizontal line indicates chance-level ($r_D=0$ or AAD accuracy = 0.5)}}  
  
  \label{fig:subject_level_recon}
\end{figure}

\begin{figure}[!t]
  \centering
  \includegraphics[width=\linewidth]{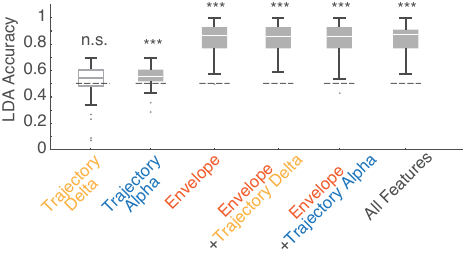}
  \caption{\textit{\textbf{Full-trial LDA decoding across feature types.}}
  \textup{LDA accuracy for envelope, trajectory (delta/alpha), and combined feature sets. Accuracy reflects the proportion of correctly classified held-out trials under leave-one-trial-out cross-validation. Each box shows the median and interquartile range across subjects. The dashed line indicates permutation-based chance level estimated separately for each feature condition (1000 label shuffles). 
  Statistical significance was assessed using one-sample right-tailed $t$-tests against subject-specific null means, with multiple comparisons corrected using the Benjamini-Hochberg FDR correction. Asterisks denote significant above-chance decoding (* $p<0.05$, ** $p<0.01$, *** $p<0.001$).}}
  \label{fig:lda_fulltrial}
\end{figure}

Stimulus reconstruction was performed using backward regularized regression with leave-one-out cross-validation to evaluate the feasibility of auditory attention decoding in the present dataset. As shown in Fig.~\ref{fig:aad_accuracy}, reconstruction performance varied across feature types and sliding-window sizes.

In the left panel, envelope exhibited a clear separation between reconstruction correlations of attended ($r_A$) and unattended ($r_U$) streams across all window sizes. The magnitude of $r_A$ for envelope remained relatively stable as window size increased, while $r_U$ remained near chance level, indicating robust attentional selectivity across window lengths. In contrast, trajectory features showed weaker correlations. The alpha trajectory feature exhibited higher $r_A$ than delta and showed a modest increase in $r_A - r_U$ separation as window size increased, while the delta trajectory remained chance level across window size.

The decoding analysis (right panel) again presented distinct feature-specific validity in AAD. Envelope-based AAD accuracy increased with window size and substantially exceeded chance even at short windows. Delta trajectory decoding remained near chance across window sizes, showing minimal improvement, while alpha trajectory decoding performed consistently above delta and exhibited modest but systematic gains with increasing window size.

\begin{figure*}[!t]
    \centering
    \includegraphics[width=1.05\linewidth]{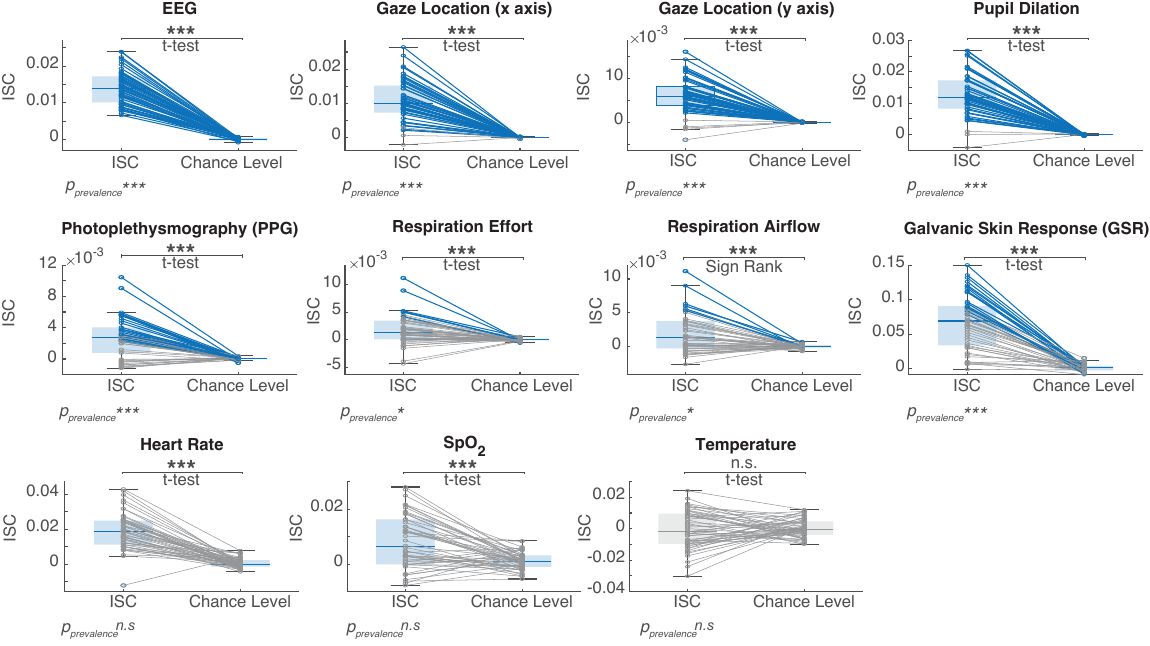}
    \caption{\textit{\textbf{Intersubject correlation across modalities.}}
    \textup{For each modality, trial-wise ISC was computed and averaged across trials to obtain one ISC value per subject. Chance level was estimated from a subject-specific permutation null distribution and summarized in the same manner.
    Boxes show the median and interquartile range across subjects; paired lines connect each subject’s ISC to its chance baseline (blue: subject-level significant difference from the null, $p<0.05$, permutation test; gray: n.s.).
    Group-level significance was assessed using a paired right-tailed test across subjects comparing ISC to chance (for each modality, a paired $t$-test was used when ISC-chance differences passed a Lilliefors normality test; otherwise a Wilcoxon signed-rank test was applied).
    To quantify prevalence, we counted the number of subjects with permutation test $p<0.05$ and evaluated whether this count exceeded the expected false-positive rate using a binomial test (reported in each panel). Statistical significance: * $p<0.05$, ** $p<0.01$, *** $p<0.001$.}}
    \label{fig:isc_modalities}
\end{figure*}

Fig.~\ref{fig:subject_level_recon} presents stimulus tracking performance at the subject level, complementing the population-level results shown above. Envelope yields consistently positive reconstruction correlations $r_D = r_A - r_U$  and higher AAD accuracy across participants, whereas trajectory features show weaker reconstruction and are only evident in a subset of subjects.


To further quantify the contribution of different feature types, we applied an LDA classifier to full-trial reconstruction correlations derived from individual features and their combinations (Fig.~\ref{fig:lda_fulltrial}). Consistent with the sliding-window analysis, envelope features yielded the highest LDA accuracy across subjects. Alpha trajectory features achieved significant above-chance performance (one-sample right-tailed $t$-test against permutation-based null means), whereas delta trajectory decoding remained at chance level and did not reach statistical significance. 

Combining envelope with trajectory features significantly improved LDA accuracy relative to trajectory-only decoding, but did not significantly exceed envelope-only decoding, consistent with prior work showing that envelope features contribute most strongly to AAD \cite{bednar_where_2020}.

These results demonstrate that the present dataset contains reliable attentional information suitable for AAD, with envelope-based features providing the dominant signal and alpha-band trajectory features offering complementary but limited support.

\subsection{Inter-subject Correlation}

ISC measures the similarity of time-varying responses across individuals exposed to the same stimulus and is widely used to characterize shared neural and physiological dynamics during naturalistic stimulation \cite{dmochowski_correlated_2012}. Here, we quantified ISC across all recorded modalities to assess signal reliability beyond EEG and to evaluate reproducible stimulus-locked temporal structure.

As shown in Fig.~\ref{fig:isc_modalities}, several modalities demonstrated above-chance ISC at the group level and in a substantial fraction of participants. In particular, EEG, gaze location, pupil dilation, and PPG exhibited consistent cross-subject temporal alignment, indicating reliable stimulus-related dynamics. In contrast, slower-varying measures such as temperature displayed minimal evidence of shared temporal structure at the trial level. Some tonic physiological signals yielded significant group-level effects but limited subject-level prevalence, such as heart rate and SpO\textsubscript{2}, suggesting heterogeneous responsiveness across participants.

Overall, these findings indicate that multiple recorded modalities contain reproducible stimulus-locked structure under naturalistic listening conditions in the present dataset. The observed ISC profiles support the suitability of these signals for downstream multimodal modeling and integration.


\section{Conclusion}
We introduced MOV-AAD, a large-scale multimodal dataset for auditory attention decoding and cognitive state assessment under realistic listening conditions. The dataset integrates EEG with synchronized physiological and behavioral recordings from 50 participants during spatially moving listening tasks.

We conducted validation analyses including stimulus reconstruction, auditory attention decoding, and intersubject correlation across modalities. The results demonstrate reliable EEG attentional modulation, cross-subject temporal alignment, and consistent multimodal physiological responses, indicating reproducible stimulus- and attention-related structure suitable for downstream modeling and benchmarking.

Combining ecological validity, multimodal coverage, and a substantial sample size, MOV-AAD supports investigations of neural and autonomic dynamics of selective attention and the development of decoding approaches in complex acoustic environments. The dataset is publicly available to facilitate reproducibility and cross-laboratory comparison.

\section{Acknowledgments}

This work was supported by the National Institute on Deafness and Other Communication Disorders (NIH-NIDCD). Additional support was provided by grants from the Marie-Josée and Henry R. Kravis Foundation.

\section{Generative AI Use Disclosure}
Generative AI tools were used only for language polishing and grammar checking. All scientific content, analyses, interpretations, and conclusions were produced and verified by the authors.

\bibliographystyle{IEEEtran}
\bibliography{merged}

\end{document}